\documentclass[12pt,titlepage]{article}
\usepackage{amsfonts}
\usepackage{amsmath}
\usepackage[dvips]{graphicx}

\begin{document}

\title{An effective singular oscillator for Duffin-Kemmer-Petiau particles
with a nonminimal vector coupling: a two-fold degeneracy}
\author{T.R. Cardoso\thanks{%
cardoso@feg.unesp.br }, L.B. Castro\thanks{%
benito@feg.unesp.br }, A.S. de Castro\thanks{%
castro@pq.cnpq.br.} \\
%EndAName
UNESP - Campus de Guaratinguet\'{a}\\
Departamento de F\'{\i}sica e Qu\'{\i}mica\\
12516-410 Guaratinguet\'{a} SP - Brazil}
\date{ }
\maketitle

\begin{abstract}
Scalar and vector bosons in the background of one-dimensional nonminimal
vector linear plus inversely linear potentials are explored in a unified way
in the context of the Duffin-Kemmer-Petiau theory. The problem is mapped
into a Sturm-Liouville problem with an effective singular oscillator. With
boundary conditions emerging from the problem, exact bound-state solutions
in the spin-0 sector are found in closed form and it is shown that the
spectrum exhibits degeneracy. It is shown that, depending on the potential
parameters, there may or may not exist bound-state solutions in the spin-1
sector. \newline


\end{abstract}

\section{Introduction}

The first-order Duffin-Kemmer-Petiau (DKP) formalism \cite{pet}-\cite{kem}
describes spin-0 and spin-1 particles and has been used to analyze
relativistic interactions of spin-0 and spin-1 hadrons with nuclei as an
alternative to their conventional second-order Klein-Gordon and Proca
counterparts. The DKP formalism enjoys a richness of couplings not capable
of being expressed in the Klein-Gordon and Proca theories \cite{gue}-\cite%
{vij}. Although the formalisms are equivalent in the case of minimally
coupled vector interactions \cite{mr}-\cite{lun}, the DKP formalism opens
news horizons as far as it allows other kinds of couplings which are not
possible in the Klein-Gordon and Proca theories. Nonminimal vector
potentials, added by other kinds of Lorentz structures, have already been
used successfully in a phenomenological context for describing the
scattering of mesons by nuclei \cite{cla1}-\cite{cla2}. Nonminimal vector
coupling with a quadratic potential \cite{Ait}, with a linear potential \cite%
{kuli}, and mixed space and time components with a step potential \cite{ccc3}%
-\cite{ccc2} and a linear potential \cite{jpa}-\cite{cascas} have been
explored in the literature. See also Ref. \cite{jpa} for a comprehensive
list of references on other sorts of couplings and functional forms for the
potential functions. In Refs. \cite{jpa}-\cite{cascas} it was shown that
when the space component of the coupling is stronger than its time component
the linear potential, a sort of vector DKP oscillator, can be used as a
model for confining bosons.

The Schr\"{o}dinger equation with a quadratic plus inversely quadratic
potential, known as singular oscillator, is an exactly solvable problem \cite%
{lan}-\cite{pr} which works for constructing solvable models of $N$
interacting bodies \cite{cal1}-\cite{cal2} as well as a basis for
perturbative expansions and variational analyses for spiked harmonic
oscillators \cite{hal1}-\cite{hal7}. Generalizations for finite-difference
relativistic quantum mechanics \cite{nag} as well as for time-dependent
parameters in the nonrelativistic version have also been considered \cite%
{cam}-\cite{dod}.

The main purpose of the present article is to report on the properties of
the DKP theory with nonminimal vector linear plus inversely linear
potentials for spin-0 and spin-1 bosons in a unified way. The problem is
mapped into an exactly solvable Sturm-Liouville problem of a Schr\"{o}%
dinger-like equation. The effective potential resulting from the mapping has
the form of the singular oscillator potential. The Schr\"{o}dinger equation
with quadratic plus inversely quadratic potential is indeed an exactly
solvable problem. Nevertheless, only positive coefficients are involved in
the well-known solution. Because we need solutions involving a repulsive as
well as an attractive inverse-square term in the effective potential, the
calculation including this generalization with proper boundary conditions at
the origin is presented. The closed form solution for the bound states is
uniquely determined with boundary conditions which emerge as a direct
consequence of the equation of motion and the normalization condition, and
do not have to be imposed. It is shown that the spectrum exhibits
degeneracy. It is also shown that the existence of bound-state solutions for
vector bosons depends on a too restrictive condition on the potential
parameters.

It should be mentioned that a somewhat less general sort of problem, only
with the space component of a nonminimal vector potential (erroneously
called pseudoscalar potential), has already appeared in the literature \cite%
{cha}.

\section{Nonminimal vector couplings in the DKP equation}

The DKP equation for a free boson is given by \cite{kem} (with units in
which $\hbar =c=1$)%
\begin{equation}
\left( i\beta ^{\mu }\partial _{\mu }-m\right) \psi =0  \label{dkp}
\end{equation}%
\noindent where the matrices $\beta ^{\mu }$\ satisfy the algebra $\beta
^{\mu }\beta ^{\nu }\beta ^{\lambda }+\beta ^{\lambda }\beta ^{\nu }\beta
^{\mu }=g^{\mu \nu }\beta ^{\lambda }+g^{\lambda \nu }\beta ^{\mu }$
\noindent and the metric tensor is $g^{\mu \nu }=\,$diag$\,(1,-1,-1,-1)$.
That algebra generates a set of 126 independent matrices whose irreducible
representations are a trivial representation, a five-dimensional
representation describing the spin-0 particles and a ten-dimensional
representation associated to spin-1 particles. The second-order Klein-Gordon
and Proca equations are obtained when one selects the spin-0 and spin-1
sectors of the DKP theory. A well-known conserved four-current is given by $%
J^{\mu }=\bar{\psi}\beta ^{\mu }\psi $\noindent $/2$ where the adjoint
spinor $\bar{\psi}$ is given by $\bar{\psi}=\psi ^{\dagger }\eta ^{0}$ with $%
\eta ^{0}=2\beta ^{0}\beta ^{0}-1$. The time component of this current is
not positive definite but it may be interpreted as a charge density. Then
the normalization condition $\int d\tau \,J^{0}=\pm 1$ can be expressed as%
\begin{equation}
\int d\tau \,\bar{\psi}\beta ^{0}\psi =\pm 2  \label{norm}
\end{equation}%
where the plus (minus) sign must be used for a positive (negative) charge.

With the introduction of nonminimal vector interactions, the DKP equation
can be written as%
\begin{equation}
\left( i\beta ^{\mu }\partial _{\mu }-m-i[P,\beta ^{\mu }]A_{\mu }\right)
\psi =0  \label{dkp2}
\end{equation}%
where $P$ is a projection operator ($P^{2}=P$ and $P^{\dagger }=P$) in such
a way that $\bar{\psi}[P,\beta ^{\mu }]\psi $ behaves like a vector under a
Lorentz transformation as does $\bar{\psi}\beta ^{\mu }\psi $. Once again $%
\partial _{\mu }J^{\mu }=0$ \cite{jpa}. If the potential is time-independent
one can write $\psi (\vec{r},t)=\phi (\vec{r})\exp (-iEt)$, where $E$ is the
energy of the boson, in such a way that the time-independent DKP equation
becomes%
\begin{equation}
\left[ \beta ^{0}E+i\beta ^{i}\partial _{i}-\left( m+i[P,\beta ^{\mu
}]A_{\mu }\right) \right] \phi =0  \label{DKP10}
\end{equation}%
In this case \ $J^{\mu }=\bar{\phi}\beta ^{\mu }\phi /2$ does not depend on
time, so that the spinor $\phi $ describes a stationary state.

The DKP equation is invariant under the parity operation, i.e. when $%
\overrightarrow{r}\rightarrow -\overrightarrow{r}$, if $\overrightarrow{A}$
changes sign, whereas $A_{0}$ remains the same. This is because the parity
operator is $\mathcal{P}=\exp (i\delta _{P})P_{0}\eta ^{0}$, where $\delta
_{P}$ is a constant phase and $P_{0}$ changes $\overrightarrow{r}$ into $-%
\overrightarrow{r}$. Because this unitary operator anticommutes with $\beta
^{i}$ and $[P,\beta ^{i}]$, they change sign under a parity transformation,
whereas $\beta ^{0}$ and $[P,\beta ^{0}]$, which commute with $\eta ^{0}$,
remain the same. Since $\delta _{P}=0$ or $\delta _{P}=\pi $, the spinor
components have definite parities. The charge-conjugation operation can be
accomplished by the transformation $\psi \rightarrow \psi _{c}=\mathcal{C}%
\psi =CK\psi $, where $K$ denotes the complex conjugation and $C$ is a
unitary matrix such that $C\beta ^{\mu }=-\beta ^{\mu }C$. Meanwhile $C$
anticommutes with $[P,\beta ^{\mu }]$. The matrix that satisfies these
relations is $C=\exp \left( i\delta _{C}\right) \eta ^{0}\eta ^{1}$. The
phase factor $\exp \left( i\delta _{C}\right) $ is equal to $\pm 1$, thus $%
E\rightarrow -E$. Note also that $J^{\mu }\rightarrow -J^{\mu }$, as should
be expected for a charge current and the charge-conjugation operation
entails no change on $A_{\mu }$. The invariance of the nonminimal vector
potential under charge conjugation means that it does not couple to the
charge of the boson. In other words, $A_{\mu }$ does not distinguish
particles from antiparticles. Hence, whether one considers spin-0 or spin-1
bosons, this sort of interaction can not exhibit Klein's paradox.

For the case of spin 0, we use the representation for the $\beta ^{\mu }$\
matrices given by \cite{ned1}%
\begin{equation}
\beta ^{0}=%
\begin{pmatrix}
\theta & \overline{0} \\
\overline{0}^{T} & \mathbf{0}%
\end{pmatrix}%
,\quad \beta ^{i}=%
\begin{pmatrix}
\widetilde{0} & \rho _{i} \\
-\rho _{i}^{T} & \mathbf{0}%
\end{pmatrix}%
,\quad i=1,2,3  \label{rep}
\end{equation}%
\noindent where%
\begin{eqnarray}
\ \theta &=&%
\begin{pmatrix}
0 & 1 \\
1 & 0%
\end{pmatrix}%
,\quad \rho _{1}=%
\begin{pmatrix}
-1 & 0 & 0 \\
0 & 0 & 0%
\end{pmatrix}
\notag \\
&&  \label{rep2} \\
\rho _{2} &=&%
\begin{pmatrix}
0 & -1 & 0 \\
0 & 0 & 0%
\end{pmatrix}%
,\quad \rho _{3}=%
\begin{pmatrix}
0 & 0 & -1 \\
0 & 0 & 0%
\end{pmatrix}
\notag
\end{eqnarray}%
\noindent $\overline{0}$, $\widetilde{0}$ and $\mathbf{0}$ are 2$\times $3, 2%
$\times $2 \ and 3$\times $3 zero matrices, respectively, while the
superscript T designates matrix transposition. Here the projection operator
can be written as \cite{gue} $P=\left( \beta ^{\mu }\beta _{\mu }-1\right)
/3=\mathrm{diag}\,(1,0,0,0,0)$. In this case $P$ picks out the first
component of the DKP spinor. The five-component spinor can be written as $%
\psi ^{T}=\left( \psi _{1},...,\psi _{5}\right) $ in such a way that the
time-independent DKP equation for a boson constrained to move along the $X$%
-axis, restricting ourselves to potentials depending only on $x$, decomposes
into%
\begin{equation*}
\left( \frac{d^{2}}{dx^{2}}+k^{2}\right) \phi _{1}=0
\end{equation*}%
\begin{equation}
\phi _{2}=\frac{1}{m}\left( E+iA_{0}\right) \,\phi _{1}  \label{dkp4}
\end{equation}%
\begin{equation*}
\phi _{3}=\frac{i}{m}\left( \frac{d}{dx}+A_{1}\right) \phi _{1},\quad \phi
_{4}=\phi _{5}=0
\end{equation*}%
where%
\begin{equation}
k^{2}=E^{2}-m^{2}+A_{0}^{2}-A_{1}^{2}+\frac{dA_{1}}{dx}  \label{k}
\end{equation}%
Meanwhile,
\begin{equation}
J^{0}=\frac{E}{m}\,|\phi _{1}|^{2},\quad J^{1}=\frac{1}{m}\text{Im}\left(
\phi _{1}^{\ast }\,\frac{d\phi _{1}}{dx}\right)  \label{corrente4}
\end{equation}

For the case of spin 1, the $\beta ^{\mu }$\ matrices are \cite{ned2}%
\begin{equation}
\beta ^{0}=%
\begin{pmatrix}
0 & \overline{0} & \overline{0} & \overline{0} \\
\overline{0}^{T} & \mathbf{0} & \mathbf{I} & \mathbf{0} \\
\overline{0}^{T} & \mathbf{I} & \mathbf{0} & \mathbf{0} \\
\overline{0}^{T} & \mathbf{0} & \mathbf{0} & \mathbf{0}%
\end{pmatrix}%
,\quad \beta ^{i}=%
\begin{pmatrix}
0 & \overline{0} & e_{i} & \overline{0} \\
\overline{0}^{T} & \mathbf{0} & \mathbf{0} & -is_{i} \\
-e_{i}^{T} & \mathbf{0} & \mathbf{0} & \mathbf{0} \\
\overline{0}^{T} & -is_{i} & \mathbf{0} & \mathbf{0}%
\end{pmatrix}
\label{betaspin1}
\end{equation}%
\noindent where $s_{i}$ are the 3$\times $3 spin-1 matrices $\left(
s_{i}\right) _{jk}=-i\varepsilon _{ijk}$, $e_{i}$ are the 1$\times $3
matrices $\left( e_{i}\right) _{1j}=\delta _{ij}$ and $\overline{0}=%
\begin{pmatrix}
0 & 0 & 0%
\end{pmatrix}%
$, while\textbf{\ }$\mathbf{I}$ and $\mathbf{0}$\textbf{\ }designate the 3$%
\times $3 unit and zero matrices, respectively. In this representation $%
P=\,\beta ^{\mu }\beta _{\mu }-2=\mathrm{diag}\,(1,1,1,1,0,0,0,0,0,0)$, i.e.
$P$ projects out the four upper components of the DKP spinor. \noindent With
the spinor written as $\psi ^{T}=\left( \psi _{1},...,\psi _{10}\right) $,
and partitioned as%
\begin{equation*}
\psi _{8}=0
\end{equation*}%
\begin{equation*}
\psi _{I}^{\left( +\right) }=\left(
\begin{array}{c}
\psi _{3} \\
\psi _{4}%
\end{array}%
\right) ,\quad \psi _{I}^{\left( -\right) }=\psi _{5}
\end{equation*}%
\begin{equation}
\psi _{II}^{\left( +\right) }=\left(
\begin{array}{c}
\psi _{6} \\
\psi _{7}%
\end{array}%
\right) ,\quad \psi _{II}^{\left( -\right) }=\psi _{2}  \label{part}
\end{equation}%
\begin{equation*}
\psi _{III}^{\left( +\right) }=\left(
\begin{array}{c}
\psi _{10} \\
-\psi _{9}%
\end{array}%
\right) ,\quad \psi _{III}^{\left( -\right) }=\psi _{1}
\end{equation*}%
the one-dimensional time-independent DKP equation can be expressed as
\begin{equation*}
\left( \frac{d^{2}}{dx^{2}}+k_{\sigma }^{2}\right) \phi _{I}^{\left( \sigma
\right) }=0
\end{equation*}%
\begin{equation}
\phi _{II}^{\left( \sigma \right) }=\frac{1}{m}\left( E+i\sigma A_{0}\right)
\,\phi _{I}^{\left( \sigma \right) }  \label{spin1-ti}
\end{equation}%
\begin{equation*}
\phi _{III}^{\left( \sigma \right) }=\frac{i}{m}\left( \frac{d}{dx}+\sigma
A_{1}\right) \phi _{I}^{\left( \sigma \right) },\quad \phi _{8}=0
\end{equation*}%
where $\sigma $ is equal to $+$ or $-$, and%
\begin{equation}
k_{\sigma }^{2}=E^{2}-m^{2}+A_{0}^{2}-A_{1}^{2}+\sigma \frac{dA_{1}}{dx}
\end{equation}%
Now the components of the four-current are%
\begin{equation}
J^{0}=\frac{E}{m}\sum\limits_{\sigma }|\phi _{I}^{\left( \sigma \right)
}|^{2},\quad J^{1}=\frac{1}{m}\text{Im}\sum\limits_{\sigma }\phi
_{I}^{\left( \sigma \right) \dagger }\,\frac{d\phi _{I}^{\left( \sigma
\right) }}{dx}  \label{CUR2}
\end{equation}%
\qquad

It is of interest to note that $\phi _{I}^{\left( +\right) }$ (in the vector
sector) obeys the same equation obeyed by the first component of the DKP
spinor in the scalar sector and, taking account of (9) and (14), $\phi _{1}$
and $\phi _{I}^{\left( \sigma \right) }$ are square-integrable functions.
Given that the interaction potentials satisfy certain conditions, we have a
well-defined Sturm-Liouville problem and hence a natural and definite method
for determining the possible discrete or continuous eigenvalues of the
system. We also note that there is only one independent component of the DKP
spinor for the spin-0 sector instead of the three required for the spin-1
sector, and that the presence of a space component of the potential might
compromise the existence of solutions for spin-1 bosons when compared to the
solutions for spin-0 bosons with the very same potentials ($A_{0}$ and $A_{1}
$). This might happen because the solution for this class of problem
consists in searching for bounded solutions for two Schr\"{o}dinger
equations. It should not be forgotten, though, that the equations for $\phi
_{I}^{\left( +\right) }$\ and $\phi _{I}^{\left( -\right) }$ are not
independent because the energy of the boson, $E$, appears in both equations.
Therefore, one has to search for bound-state solutions for $\phi
_{I}^{\left( +\right) }$\ and $\phi _{I}^{\left( -\right) }$ with a common
energy. This amounts to say that the solutions for the spin-1 sector of the
DKP theory can be obtained from a restricted class of solutions of the
spin-0 sector.

\section{The linear plus inversely linear potential}

Now we are in a position to use the DKP equation with specific forms for
vector interactions. Let us focus our attention on potentials in the linear
plus inversely linear form, viz.
\begin{equation}
A_{0}=m^{2}\omega _{0}|x|+\frac{g_{0}}{|x|},\quad A_{1}=m^{2}\omega _{1}x+%
\frac{g_{1}}{x}  \label{sc1a}
\end{equation}%
where the coupling constants, $\omega _{0},\omega _{1},g_{0}$ and $g_{1}$,
are real dimensionless parameters. Our problem is to solve (\ref{dkp4}) and (%
\ref{spin1-ti}) for $\phi _{1}$ (in the scalar sector) and $\phi
_{I}^{\left( \sigma \right) }$ (in the vector sector), and to determine the
allowed energies. Although the absolute value of $x$ in $A_{0}$ is
irrelevant in the effective equations for $\phi _{1}$ and $\phi _{I}^{\left(
\sigma \right) }$, it is there for ensuring the covariance of the DKP theory
under parity. It follows that the DKP spinor will have a definite parity and
$A^{\mu }$ will be a genuine four-vector. In this case the first equations
of (\ref{dkp4}) and (\ref{spin1-ti}) transmutes into%
\begin{equation}
H_{\sigma }\,\Phi _{\sigma }=\varepsilon _{\sigma }\Phi _{\sigma }
\label{sc1b}
\end{equation}%
where $\Phi _{\sigma }$ is equal to $\phi _{1}$ for the scalar sector, and
to $\phi _{I}^{\left( \sigma \right) }$ for the vector sector, with

\begin{equation}
H_{\sigma }=-\frac{1}{2m}\frac{d^{2}}{dx^{2}}+V_{\sigma }\,  \label{Heff}
\end{equation}%
\begin{equation}
\varepsilon _{\sigma }=\frac{E^{2}-m^{2}+m^{2}\left[ \sigma \omega
_{1}+2\left( \omega _{0}g_{0}-\omega _{1}g_{1}\right) \right] }{2m}
\label{sc2}
\end{equation}%
and%
\begin{equation}
V_{\sigma }=\frac{1}{2}m\Omega ^{2}x^{2}+\frac{\alpha _{\sigma }}{x^{2}}
\label{effpot}
\end{equation}%
with
\begin{equation}
\Omega ^{2}=m^{2}\left( \omega _{1}^{2}-\omega _{0}^{2}\right) ,\quad \alpha
_{\sigma }=\frac{g_{1}\left( g_{1}+\sigma \right) -g_{0}^{2}}{2m}
\label{sc3}
\end{equation}

\noindent The set (\ref{sc1b})-(\ref{sc3}), with $\alpha _{\sigma }=0$ ($%
\alpha _{\sigma }\neq 0$) and $\Omega ^{2}>0$, is precisely the Schr\"{o}%
dinger equation for the nonrelativistic nonsingular (singular) harmonic
oscillator. For $\alpha _{\sigma }<0$ and $\Omega ^{2}=0$ the effective
potential has also a form that would make allowance for bound-state
solutions with $\varepsilon _{\sigma }<0$. All the remaining possibilities
for $\alpha _{\sigma }$ and $\Omega ^{2}$ make the effective potential
either everywhere repulsive or there appears an attractive region about $x=0$
between potential barriers ($\alpha _{\sigma }<0$ and $\Omega ^{2}<0$) which
lead to boson tunneling. It is well known that the singularity at $x=0$ for $%
\alpha _{\sigma }<0$ menaces the boson to collapse to the center \cite{lan}
so that the condition $\alpha _{\sigma }\geq \alpha _{\mathtt{crit}%
}=-1/\left( 8m\right) $ (with $\Omega ^{2}>0$) is required for the formation
of bound-state solutions. We shall limit ourselves to study the bound-state
solutions.

Since $V_{\sigma }$ is invariant under reflection through the origin ($%
x\rightarrow -x$), eigenfunctions with well-defined parities can be built.
Thus, one can concentrate attention on the positive half-line and impose
boundary conditions on $\Phi _{\sigma }$ at $x=0$ and $x=\infty $.
Continuous eigenfunctions on the whole line with well-defined parities can
be constructed by taking symmetric and antisymmetric linear combinations of $%
\Phi _{\sigma }$. These new eigenfunctions possess the same eigenenergy,
then, in principle, there is a two-fold degeneracy. As $x\rightarrow 0$, the
solution behaves as $Cx^{s_{\sigma }}$, where $C\neq 0$ is a constant and $%
s_{\sigma }$ is a solution of the algebraic equation

\begin{equation}
s_{\sigma }(s_{\sigma }-1)-2m\alpha _{\sigma }=0  \label{17}
\end{equation}%
viz.

\begin{equation}
s_{\sigma }=\frac{1}{2}\left( 1\pm \sqrt{1+8m\alpha _{\sigma }}\right)
\label{18}
\end{equation}

\noindent where $s_{\sigma }$ is not necessarily a real quantity.
Nevertheless, $J^{1}$ for a stationary state, as expressed by (\ref%
{corrente4}) and (\ref{CUR2}), is the same at all points of the $X$-axis and
vanishes for a bound-state solution (because $\Phi _{\sigma }$ vanishes as $%
x\rightarrow \infty $), so we demand that $s_{\sigma }\in
%TCIMACRO{\U{211d} }%
%BeginExpansion
\mathbb{R}
%EndExpansion
$ and so $\alpha _{\sigma }\geq \alpha _{\mathtt{crit}}$. Normalizability of
$\Phi _{\sigma }$ also requires $s_{\sigma }>-1/2$ and due to the two-fold
possibility of values of $s_{\sigma }$ for $\alpha _{\mathtt{crit}}<\alpha
_{\sigma }<3/(8m)$, it seems that the solution of our problem can not be
uniquely determined. This ambiguity can be overcome by a regularization of
the potential. Following the steps of Ref. \cite{lan}, we replace $V_{\sigma
}(x)$ by $V_{\sigma }(x_{0})$ \ for $x<x_{0}\approx 0$ and after using the
continuity conditions for $\Phi _{\sigma }$ and $d\Phi _{\sigma }/dx$ in the
cutoff \ we take the limit $x_{0}\rightarrow 0$. It turns out that the
solution with the lesser value of $s_{\sigma }$ is suppressed relative to
that one involving the greater value as $x_{0}\rightarrow 0$. Thus, the
minus sign in (\ref{18}) must be ruled out for $\alpha _{\sigma }\neq 0$ in
such a way that $\Phi _{\sigma }(0)=0$. Hence, $s_{\sigma }=0$ or $s_{\sigma
}\geq 1/2$. The homogeneous Dirichlet boundary condition ($\Phi _{\sigma
}\left( 0\right) =0$) is essential whenever $\alpha _{\sigma }\neq 0$,
nevertheless it also develops for $\alpha _{\sigma }=0$ when $s_{\sigma }=1$
but not for $s_{\sigma }=0$. The continuity of $\Phi _{\sigma }$ at the
origin excludes the possibility of an odd-parity eigenfunction for $s=0$,
and effects on the even-parity eigenfunction for $s=1$ are due to the
continuity of its first derivative. Effects due to the potential on $d\Phi
_{\sigma }/dx$ in the neighbourhood of the origin can be evaluated by
integrating (\ref{sc1b}) from $-\delta $ to $+\delta $ and taking the limit $%
\delta \rightarrow 0$. The connection formula between $d\Phi _{\sigma }/dx$
at the right and $d\Phi _{\sigma }/dx$ at the left can be summarized as%
\begin{equation}
\lim_{\delta \rightarrow 0}\left. \frac{d\Phi _{\sigma }}{dx}\right\vert
_{x=-\delta }^{x=+\delta }=2m\alpha _{\sigma }\lim_{\delta \longrightarrow
0}\int_{-\delta }^{+\delta }dx\;\frac{\Phi _{\sigma }}{x^{2}}
\end{equation}%
so that%
\begin{equation}
\lim_{\delta \rightarrow 0}\left. \frac{d\Phi _{\sigma }}{dx}\right\vert
_{x=-\delta }^{x=+\delta }=\left\{
\begin{array}{c}
2mC\alpha _{\sigma }\left. \frac{|x|^{s_{\sigma }-1}}{s_{\sigma }-1}%
\right\vert _{x=-\delta }^{x=+\delta }\text{, for }\Phi _{\sigma }\text{
symmetric with }\alpha _{\sigma }\neq 0 \\
\\
0\text{, otherwise}%
\end{array}%
\right.
\end{equation}%
Therefore, the eigenfunctions have a first derivative continuous at the
origin, except for $s_{\sigma }<1$ when the eigenfunction is not
differentiable at the origin. Thus, the bound-state solutions for the
singular oscillator potential are two-fold degenerate: for a same energy
there is an odd-parity solution with a continuous first derivative, and an
even-parity solution with a continuous first derivative if $s_{\sigma }>1$
or a discontinuous first derivative if $s_{\sigma }<1$. There is no
odd-parity solution for $s_{\sigma }=0$ and there develops the homogeneous
Neumann condition ($\left. d\Phi _{\sigma }/dx\right\vert _{x=0}=0$), and
there is no even-parity solution for $s_{\sigma }=1$. Thus, the bound-state
solutions for the nonsingular oscillator potential are nondegenerate.

We shall now consider separately two different possibilities for $\Omega $,
namely $\Omega =0$ and $\Omega \neq 0$.

\subsection{$\Omega =0$}

Defining%
\begin{equation}
z_{\sigma }=2\sqrt{-2m\varepsilon _{\sigma }}\,x  \label{z}
\end{equation}%
\noindent for $x>0$, and using the set (\ref{sc1b})-(\ref{sc3}), one obtains
a special case of Whittaker%
%TCIMACRO{\U{b4}}%
%BeginExpansion
\'{}%
%EndExpansion
s differential equation \cite{abr}%
\begin{equation}
\frac{d^{2}\Phi _{\sigma }}{dz_{\sigma }^{2}}+\left( -\frac{1}{4}-\frac{%
2m\alpha _{\sigma }}{z_{\sigma }^{2}}\right) \Phi _{\sigma }=0
\end{equation}%
The normalizable asymptotic form of the solution as $z_{\sigma }\rightarrow
\infty $ is $e^{-z_{\sigma }/2}$ with $z_{\sigma }>0$. Notice that this
asymptotic behaviour rules out the possibility $\varepsilon _{\sigma }>0$,
as has been pointed out already based on qualitative arguments. The exact
solution can now be written as%
\begin{equation}
\Phi _{\sigma }=z_{\sigma }^{s_{\sigma }}w(z_{\sigma })e^{-z_{\sigma }/2}
\label{comb1}
\end{equation}%
where $w$ is a regular solution of the confluent hypergeometric equation
(Kummer's equation) \cite{abr}
\begin{equation}
z_{\sigma }\frac{d^{2}w}{dz_{\sigma }^{2}}+(b_{\sigma }-z_{\sigma })\frac{dw%
}{dz_{\sigma }}-a_{\sigma }w=0  \label{35}
\end{equation}%
\noindent with the definitions%
\begin{equation}
a_{\sigma }=s_{\sigma },\quad b_{\sigma }=2s_{\sigma }  \label{36}
\end{equation}%
The general solution of \ (\ref{35}) is expressed in terms of the confluent
hypergeometric functions (Kummer's functions) $_{1}F_{1}(a_{\sigma
},b_{\sigma },z_{\sigma })$ (or $M(a_{\sigma },b_{\sigma },z_{\sigma })$)
and $_{2}F_{0}(a_{\sigma },1+a_{\sigma }-b_{\sigma },-1/z_{\sigma })$ (or $%
U(a_{\sigma },b_{\sigma },z_{\sigma })$):
\begin{equation}
w=A_{\sigma }\,\,_{1}F_{1}(a_{\sigma },b_{\sigma },z_{\sigma })+B_{\sigma
}z_{\sigma }^{-a_{\sigma }}\,_{2}F_{0}(a_{\sigma },1+a_{\sigma }-b_{\sigma
},-\frac{1}{z_{\sigma }}),\quad b_{\sigma }\neq -\tilde{n}_{\sigma }
\label{W}
\end{equation}%
where $\tilde{n}_{\sigma }$ is a nonnegative integer. Due to the singularity
of the second term at $z_{\sigma }=0$, only choosing $B_{\sigma }=0$ gives a
behavior at the origin which can lead to square-integrable solutions.
Furthermore, the requirement of finiteness for $\Phi _{\sigma }$ at $%
z_{\sigma }=\infty $ implies that the remaining confluent hypergeometric
function ($_{1}F_{1}(a_{\sigma },b_{\sigma },z_{\sigma })$) should be a
polynomial. This is because $_{1}F_{1}(a_{\sigma },b_{\sigma },z_{\sigma })$
goes as $e^{z_{\sigma }}$ as $z_{\sigma }$ goes to infinity unless the
series breaks off. This demands that $a_{\sigma }=-n_{\sigma }$, where $%
n_{\sigma }$ is also a nonnegative integer. This requirement combined with (%
\ref{36}) implies that the existence of bound-state solutions for pure
inversely quadratic potentials ($|\omega _{1}|=|\omega _{0}|$) is out of
question.

\subsection{$\Omega \neq 0$}

As for $\Omega \neq 0$, it is convenient to define the dimensionless
quantity $\xi $,

\negthinspace
\begin{equation}
\xi =m\sqrt{\Omega ^{2}}\,x^{2}  \label{15}
\end{equation}%
\noindent \noindent and using (\ref{sc1b})-(\ref{sc3}), one obtains the
complete form for Whittaker%
%TCIMACRO{\U{b4}}%
%BeginExpansion
\'{}%
%EndExpansion
s equation \cite{abr}

\begin{equation}
\,\xi \frac{d^{2}\Phi _{\sigma }}{d\xi ^{2}}+\frac{1}{2}\frac{d\Phi _{\sigma
}}{d\xi }+\left( \frac{\varepsilon _{\sigma }}{2\sqrt{\Omega ^{2}}}-\frac{%
\xi }{4}-\frac{m\alpha _{\sigma }}{2\xi }\right) \Phi _{\sigma }=0
\label{16}
\end{equation}

\noindent The normalizable asymptotic form of the solution as $\xi
\rightarrow \infty $ is given by $e^{-\xi /2}$ only for $\Omega ^{2}>0$ ($%
|\omega _{1}|>|\omega _{0}|$). The solution for all $\xi $ can now be
written as%
\begin{equation}
\Phi _{\sigma }=\xi ^{s_{\sigma }/2}w(\xi )e^{-\xi /2}  \label{comb2}
\end{equation}%
where $w(\xi )$ is a regular solution from
\begin{equation}
\xi \frac{d^{2}w}{d\xi ^{2}}+(b_{\sigma }-\xi )\frac{dw}{d\xi }-a_{\sigma
}w=0
\end{equation}%
with%
\begin{eqnarray}
a_{\sigma } &=&\frac{b_{\sigma }}{2}-\frac{\varepsilon _{\sigma }}{2|\Omega |%
}  \notag \\
&&  \label{036} \\
b_{\sigma } &=&s_{\sigma }+1/2  \notag
\end{eqnarray}%
\noindent Then, $w$ is expressed as $_{1}F_{1}(a_{\sigma },b_{\sigma },\xi )$
and in order to furnish normalizable $\Phi _{\sigma }$, the confluent
hypergeometric function must be a polynomial. This demands that $a_{\sigma
}=-n_{\sigma }$ and $b_{\sigma }\neq -\tilde{n}_{\sigma }$. Note that $%
_{1}F_{1}(-n_{\sigma },b_{\sigma },\xi )$ is proportional to the generalized
Laguerre polynomial $L_{n_{\sigma }}^{\left( b_{\sigma }-1\right) }(\xi )$,
a polynomial of degree $n_{\sigma }$. The requirement $a_{\sigma
}=-n_{\sigma }$ combined with the top line of (\ref{036}), also implies into
quantized effective eigenvalues:
\begin{equation}
\varepsilon _{\sigma }=\left( 2n_{\sigma }+s_{\sigma }+\frac{1}{2}\right)
|\Omega |,\qquad n_{\sigma }=0,1,2,\ldots   \label{37}
\end{equation}%
\noindent with
\begin{equation}
\Phi _{\sigma }\propto \xi ^{s_{\sigma }/2}e_{\;}^{-\xi /2}\;L_{n_{\sigma
}}^{\left( s_{\sigma }-1/2\right) }\left( \xi \right)   \label{38}
\end{equation}%
\noindent\

In summary, only for $\Omega ^{2}>0$ ($\varepsilon _{\sigma }>0$) and $%
\alpha _{\sigma }\geq \alpha _{\mathtt{crit}}$ the potentials (\ref{sc1a})
are able to furnish bound states. The spectrum is purely discrete and the
solution is expressed in an exact closed form. It is instructive to note
that $s_{\sigma }=0$ or $s_{\sigma }=1$ for the case $\alpha _{\sigma }=0$
and the associated Laguerre polynomial $L_{n_{\sigma }}^{\left( -1/2\right)
}(\xi )$ and $L_{n_{\sigma }}^{\left( +1/2\right) }(\xi )$ are proportional
to $H_{2n_{\sigma }}\left( \sqrt{\xi }\right) $ or $\xi ^{-1/2}H_{2n_{\sigma
}+1}\left( \sqrt{\xi }\right) $, respectively \cite{abr}. Therefore, the
solution for the effective nonsingular harmonic oscillator can be succinctly
written in the customary form in terms of Hermite polynomials:%
\begin{equation}
\varepsilon _{\sigma }=\left( n_{\sigma }+\frac{1}{2}\right) |\Omega
|,\qquad n_{\sigma }=0,1,2,\ldots  \label{410a}
\end{equation}%
with $\Phi _{\sigma }$ proportional to
\begin{equation}
e^{-\xi /2}H_{n_{\sigma }}\left( \sqrt{\xi }\right)  \label{410b}
\end{equation}

The preceding analyses shows that the effective eigenvalues for the scalar
sector are equally spaced with a step given by $2|\Omega |$ when the
effective potential is singular at the origin. It is remarkable that the
level stepping is independent of the sign and intensity of the parameter
responsible for the singularity of the potential. We also note that the
effective spectrum varies continuously with $\alpha _{+}$ as far as $\alpha
_{+}\neq 0$. When the effective potential becomes nonsingular ($\alpha
_{+}=0 $) the step switches abruptly to $|\Omega |$. There is a clear phase
transition when $\alpha _{+}\rightarrow 0$ due the disappearance of the
singularity. In the limit as $\alpha _{\sigma }\rightarrow 0$ the Neumann
boundary condition, in addition to the Dirichlet boundary condition always
present for $\alpha _{+}\neq 0$, comes to the scene. This occurrence permits
the appearance of even Hermite polynomials and their related eigenvalues,
which intercalate among the pre-existent eigenvalues related to odd Hermite
polynomials. The appearance of even Hermite polynomials makes $\Phi
_{+}(0)\neq 0$ and this boundary condition is never permitted when the
singular potential is present, even though $\alpha _{+}$ can be small. You
might also understand the lack of such a smooth transition by starting from
a nonsingular potential ($\alpha _{+}=0$), when the solution of the problem
involves even and odd Hermite polynomials, and then turning on the singular
potential as a perturbation of the nonsingular potential. Now the
\textquotedblleft perturbative singular potential\textquotedblright\ by
nature demands, if is either attractive or repulsive, that $\Phi _{+}(0)=0$
so that it naturally kills the solution involving even Hermite polynomials.
Furthermore, there is no degeneracy in the spectrum for $\alpha _{+}=0$.

Now we move on to match a common energy to the spin-1 boson problem. The
compatibility of the solutions for $\Phi _{+}=\phi _{I}^{\left( +\right) }$
and $\Phi _{-}=\phi _{I}^{\left( -\right) }$ demands that the quantum
numbers $n_{+}$ and $n_{-}$ must satisfy the relation%
\begin{equation}
n_{+}-n_{-}=R(\omega _{0},\omega _{1})
\end{equation}%
for the nonsingular potential, and%
\begin{equation}
n_{+}-n_{-}=\frac{1}{2}\left[ R(\omega _{0},\omega _{1})-\sqrt{1+8m\alpha
_{+}}+\sqrt{1+8m\alpha _{-}}\right]
\end{equation}%
for the singular potential. Here%
\begin{equation*}
R(\omega _{0},\omega _{1})=\frac{\mathrm{sgn}(\omega _{1})}{\sqrt{1-\left(
\frac{\omega _{0}}{\omega _{1}}\right) ^{2}}}
\end{equation*}%
and $\mathrm{sgn}(\omega _{1})$ stands for the sign function. For the
nonsingular potential $|n_{+}-n_{-}|$ varies from 1 to infinity (as $|\omega
_{0}/\omega _{1}|$ varies from 0 to 1), whereas for the singular potential $%
|n_{+}-n_{-}|$ varies from the minor to the major value between $\left\vert
R(\omega _{0},\omega _{1})\right\vert /2$ and $\left\vert R(\omega
_{0},\omega _{1})\pm \sqrt{8g_{1}}\right\vert /2$ (the $\pm $ sign
corresponding to $\alpha _{\pm }=\alpha _{\mathtt{crit}}$). These
constraints on the nodal structure of $\phi _{I}^{\left( +\right) }$ and $%
\phi _{I}^{\left( -\right) }$ dictate that acceptable solutions only occur
for a restricted number of possibilities for the potential parameters and
that no solution should be expected for $g_{1}<0$. More than this, these
constraints exclude a few low-lying quantum numbers for a given set of
possible solutions for the simple reason that $n_{+}$ and $n_{-}$ are
nonnegative integers. In conclusion, the possible bound-state solutions for
spin-1 bosons exist only for a finite set of potential parameters.

\begin{figure}[th]
\begin{center}
\includegraphics[width=9cm, angle=0]{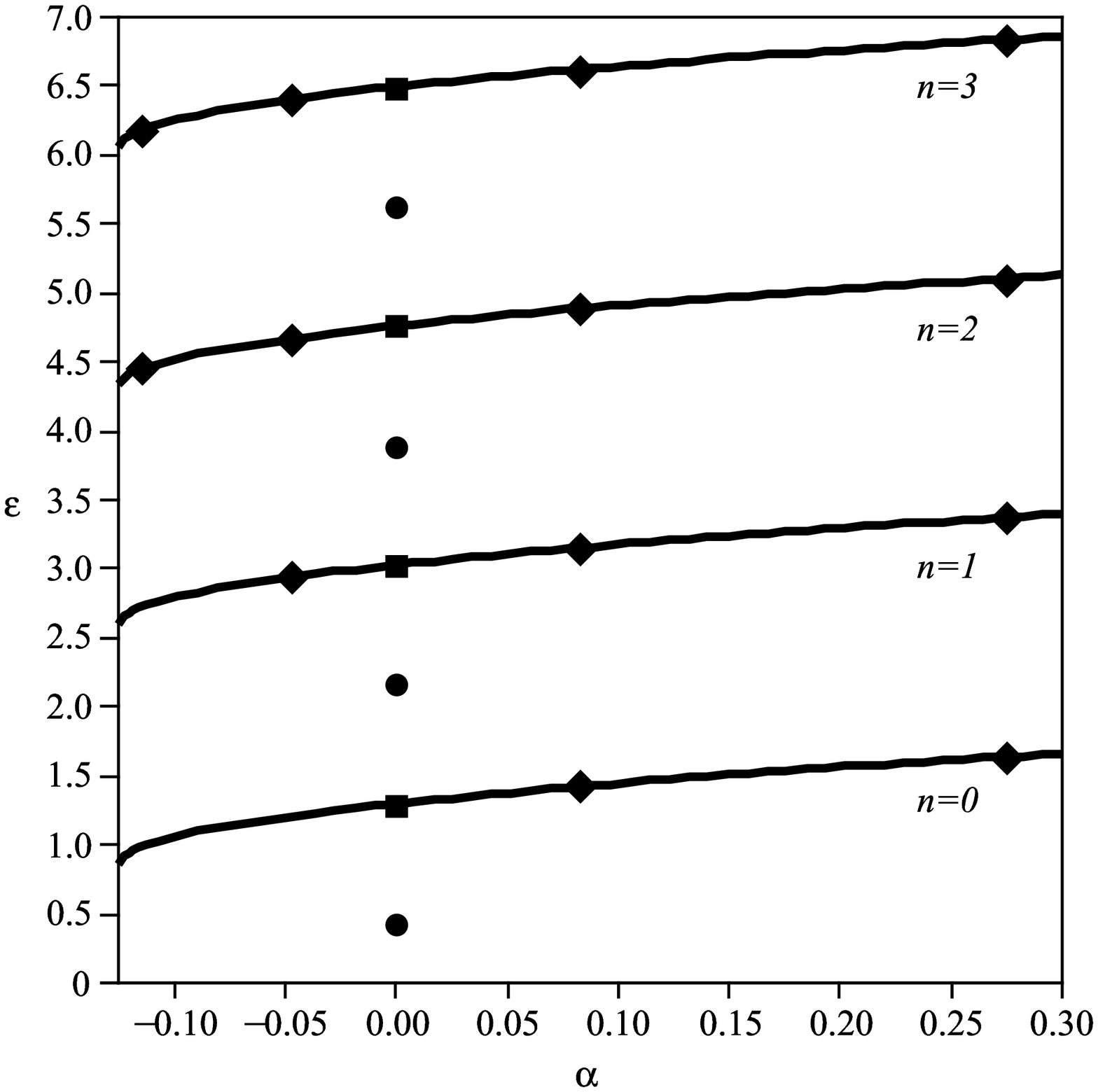}
\end{center}
\par
\vspace*{-0.1cm}
\caption{Effective spectrum ($\protect\varepsilon =\protect\varepsilon %
_{+}/m $ and $n=n_{+}$) for spin-0 bosons as a function of $\protect\alpha =%
\protect\alpha _{+}\neq 0$ (full line) with $\protect\omega _{1}=m=1$, $%
\protect\omega _{0}=1/2$ and $g_{0}=0$. The circles and boxes stand for even
and odd solutions for the case $\protect\alpha _{+}=0$, respectively. The
effective spectrum for spin-1 bosons are represented by the diamond symbols
for $n_{+}-n_{-}=+2,+1,0,-1$ corresponding to $\protect\alpha =-0.115$, $%
-0.047$, $0.083$, $0.275$ respectively.}
\label{Fig1}
\end{figure}

To have a better understanding of the spectrum, we plot in Figure \ref{Fig1}
the effective spectrum for the four low-lying levels as a function of $%
\alpha _{+}$, for fixed $\omega _{0}$ and $g_{0}$. From this figure we see
the phase transition for the spin-0 spectrum at $\alpha _{+}=0$ and the
severe restriction on the solutions for the spin-1 spectrum. Notice that no
bound-state solution for the spin-1 spectrum exists if $\alpha _{+}$ exceeds
$|R(\omega _{0},\omega _{1})|/2$ and that not all the bound states have a
nodeless $\Phi _{+}$ for the ground state.

\section{Conclusions}

We approached the problem of a particle in a (3+1)-dimensional world despite
the restriction to the one-dimensional movement. The motion on axis allow us
to explore the physical consequences of the nonminimal vector coupling in a
mathematically simpler and more physically transparent way. There is no
angular momentum so that there is no spin-orbit coupling. The more general
potential matrix ensuring a conserved four-current can be written in terms
of well-defined Lorentz structures. For the spin-0 sector there are two
scalar, two vector and two tensor terms \cite{gue}, whereas for the spin-1
sector there are two scalar, two vector, a pseudoscalar, two pseudovector
and eight tensor terms \cite{vij}. There is no pseudoscalar potential in the
spin-0 sector of the DKP theory. In fact, the space component of a
nonminimal vector potential is used in Ref. \cite{cha}. We described spin-0
particles by a five-component spinor, of which only one component is
independent instead the two required in Ref. \cite{cha}. Also, the number of
degrees of freedom described by the ten-component spinor in the spin-1
sector reduces correctly to three. By considering space plus time components
of a nonminimal vector potential, we showed that scalar and vector bosons in
the background of one-dimensional linear plus inversely linear potential is
mapped into a Sturm-Liouville problem with an effective singular oscillator.
The existence of solutions for spin-0 bosons requires $|\omega _{1}|>|\omega
_{0}|$ and $g_{1}\left( g_{1}+\sigma \right) -g_{0}^{2}\geq -1/4$. In words,
the time component of the nonminimal vector potential ($A_{0}$) alone can
not hold bound states, and the presence of the time component in the
inversely linear potential lessens the interval of parameters which provides
confinement. As for spin-1 bosons, confining solutions requires $g_{1}>0$
and when compared to the solutions for spin-0 bosons with the very same
potentials ($A_{0}$ and $A_{1}$) they can only be obtained from a restricted
class of solutions of the spin-0 sector. It may be salutary to note that,
because the differential equation (\ref{sc1b}) has a singularity at the
origin for $\alpha _{\sigma }\neq 0$, one could expect the existence of
singular solutions for $\Phi _{\sigma }$ so that we are not free to suppose
that $s_{\sigma }>0$. Due to the fact that $\Phi _{\sigma }$, whether they
are even or odd, disappears at the origin for $\alpha _{\sigma }\neq 0$ this
one-dimensional quantum-mechanical problem is degenerate and $H_{\sigma }$
in (\ref{Heff}) is a self-adjoint operator.

\bigskip \bigskip\bigskip \bigskip

\noindent {\textbf{Acknowledgments}}

This work was supported in part by means of funds provided by Coordena\c{c}%
\~{a}o de Aperfei\c{c}oamento de Pessoal de N\'{\i}vel Superior\ (CAPES) and
Conselho Nacional de Desenvolvimento Cient\'{\i}fico e Tecnol\'{o}gico
(CNPq). The authors gratefully acknowledge the anonymous referees for their
valuable comments and suggestions.

\newpage

\end{document}